\documentclass[sigconf, nonacm]{acmart}
\usepackage{tikz}
\usepackage{subcaption}
\usepackage{float}
\usepackage[T1]{fontenc}
\usepackage[utf8]{inputenc}

\usetikzlibrary{positioning,arrows.meta}

\begin{document}
\title{A Demonstration of SQLyzr: A Platform for Fine-Grained Text-to-SQL Evaluation and Analysis}
\newcommand{\name}{SQLyzr}
\newcommand{\sys}{platform}
\newcommand{\Sys}{Platform}
\newcommand{\ts}{text-to-SQL}
\newcommand{\Ts}{Text-to-SQL}
\newcommand{\mdl}{model}
\newcommand{\Mdl}{Model}
\newcommand{\tss}{text-to-SQL \mdl}
\newcommand{\Tss}{Text-to-SQL \mdl}
\newcommand{\signpost}[1]{\noindent\textbf{#1}.}
\newcommand{\sug}[1]{{\color{cyan}{#1}}}
\newcommand{\din}{DIN-SQL}
\newcommand{\dail}{DAIL-SQL}
\newcommand{\spider}{Spider}
\newcommand{\bird}{BIRD}
\newcommand{\beaver}{BEAVER}
\newcommand{\cat}[1]{c#1}
\newcommand{\ea}{Execution Accuracy}
\newcommand{\exm}{Exact Match}
\newcommand{\rea}{\lowercase{\REA}}
\newcommand{\gpt}{ChatGPT}
\newcommand{\et}{Execution Time}
\newcommand{\etc}{Execution Time Consistency}
\newcommand{\tu}{Token Usage}
\newcommand{\cc}{Complexity Consistency}
\newcommand{\icl}{In Context Learning}
\newcommand{\gptfour}{GPT4}
\newcommand{\sqlite}{SQLite}
\newcommand{\oai}{OpenAI}
\newcommand{\sub}{subcategory}
\newcommand{\subs}{subcategories}
\newcommand{\Sub}{Subcategory}
\newcommand{\Subs}{Subcategories}
\newcommand{\gold}{ground truth}
\newcommand{\Gold}{Ground truth}
\newcommand{\fig}{diagram}
\newcommand{\gmin}{GPT-4o-mini}
\newcommand{\tsp}{question-SQL}
\newcommand{\Tsp}{Question-SQL}
\newcommand{\sqs}{SQLShare}
\newcommand{\wl}{{workload}}
\newcommand{\Wl}{{Workload}}
\newcommand{\ds}{{dataset}}
\newcommand{\Ds}{{Dataset}}
\newcommand{\w}{{data point}}
\newcommand{\sref}[1]{Section \ref{#1}}
\newcommand{\aref}[1]{Appendix \ref{#1}}
\newcommand{\sepid}[1]{{\footnotesize\color{red}[#1]}}
\newcommand{\cor}{{correctness}}
\newcommand{\Cor}{{Correctness}}

\newcommand{\Exec}{Exec}
\newcommand{\EA}{EA}
\newcommand{\EM}{EM}
\newcommand{\Norm}{Norm}
\newcommand{\Eq}{Eq}
\newcommand{\Time}{Time}
\newcommand{\QET}{QET}
\newcommand{\median}{meidan}
\newcommand{\Tok}{Tok}
\newcommand{\TU}{TU}
\newcommand{\Cat}{Cat}
\newcommand{\CC}{CC}
\newcommand{\ETC}{ETC}
\newcommand{\Metric}{Metric}
\newcommand{\REA}{REA}
\newcommand{\ghurl}{https://github.com/sepideh-abedini/SQLyzr}

\newcommand{\catcount}{882}
\newcommand{\q}[1]{``#1''}

\definecolor{sqlcolor}{RGB}{0, 51, 153}
\newcommand{\sql}[1]{\texttt{\textbf{\textcolor{sqlcolor}{#1}}}}

\emergencystretch 2em

\author{Sepideh Abedini}
\affiliation{%
  \institution{University of Waterloo}
}
\email{sepideh.abedini@uwaterloo.ca}

\author{M. Tamer Özsu}
\affiliation{%
  \institution{University of Waterloo}
}
\email{tamer.ozsu@uwaterloo.ca}

\begin{abstract}
\Tss s have significantly improved with the adoption of Large Language Models (LLMs), leading to their increasing use in real-world applications. Although many benchmarks exist for evaluating the performance of \tss s, they often rely on a single aggregate score, lack evaluation under realistic settings, and provide limited insight into model behaviour across different query types. In this work, we present \name, a comprehensive benchmark and evaluation platform for \tss s. \name\ incorporates a diverse set of evaluation metrics that capture multiple aspects of generated queries, while enabling more realistic evaluation through workload alignment with real-world SQL usage patterns and database scaling. It further supports fine-grained query classification, error analysis, and \wl\ augmentation, allowing users to better diagnose and improve \tss s. This demonstration showcases these capabilities through an interactive experience. Through \name's graphical interface, users can customize evaluation settings, analyze fine-grained reports, and explore additional features of the platform. We envision that \name\ facilitates the evaluation and iterative improvement of \tss s by addressing key limitations of existing benchmarks.
The source code of \name\ is available at \url{\ghurl}. 
\end{abstract}

\maketitle

\section{Introduction}
\label{sec:intro}

Relational database management systems (RDBMSs) are widely used to store and manage structured data across many domains, such as healthcare, finance, and enterprise systems \cite{around}. Accessing this data typically requires knowledge of SQL, which can be challenging for non-expert users. \Tss s aim to address this challenge by translating natural language utterances into executable SQL queries \cite{era-survey}, enabling intuitive access to structured data. Reducing barriers to accessing RDBMSs using natural language has long been a goal, going back to early work by Codd~\cite{Codd:1974aa}, followed by many subsequent efforts, e.g., \cite{Hendrix:1978aa, Popescu:2003aa, Zhong:2017ab, Wang:2020aa}. 

Current state-of-the-art approaches rely on large language models (LLMs) to generate SQL queries from natural language questions \cite{llmsurvey}. Recent advances in LLMs have significantly improved the performance of \tss s. However, existing benchmarks have not kept pace with these advancements and suffer from several limitations that restrict their effectiveness for evaluating \tss s in real-world deployment settings.

First, existing benchmarks typically report only a single aggregate correctness score, which does not reveal which query types are more challenging for a model. Moreover, relying solely on correctness overlooks other important aspects of the generated queries, such as execution efficiency and structural complexity, which are critical factors for deploying \tss s in production environments.

Second, these benchmarks rely on fixed and small-scale databases. While this simplifies the evaluation, it does not capture how generated queries behave under realistic, large-scale settings, particularly in terms of efficiency.

Third, these benchmarks often use workloads that do not reflect real-world SQL usage patterns, limiting their ability to reliably predict model performance in practical deployments.

Finally, traditional benchmarks are inherently static. Once a \mdl\ achieves a relatively high score, subsequent evaluations provide limited diagnostic value. As a result, developers need to construct new test cases or switch benchmarks to identify remaining weaknesses. Consequently, existing benchmarks are often used as static, one-time evaluation tools, and are not well suited for integration into iterative \mdl\ development workflows.

In this work, we demonstrate \name~\cite{abedini2026sqlyzr}, a comprehensive benchmark and evaluation platform for \tss s designed to address these limitations. \name\ defines a benchmark specification consisting of a workload, a dataset, and a set of evaluation metrics. In addition, it provides a configurable benchmarking platform that supports flexible evaluation settings and enables fine-grained analysis, workload augmentation, and evaluation under scaled database settings. While the benchmark specification and evaluation framework are described in~\cite{abedini2026sqlyzr}, this work focuses on the practical usage and analysis capabilities of the platform through an interactive demonstration, where users can configure evaluations and explore \name's key features via its graphical user interface (GUI).

\section{\name\ Overview}
\label{sec:overview}
\name\ consists of two complementary components: (i) a benchmark specification that defines a workload, a dataset, and a set of evaluation metrics (\sref{sec:benchmark}), and (ii) a benchmarking platform that enables configurable evaluation of one or more \tss s (\sref{sec:framework}).

\begin{figure}[t!]
    \centering
    \includegraphics[width=\linewidth]{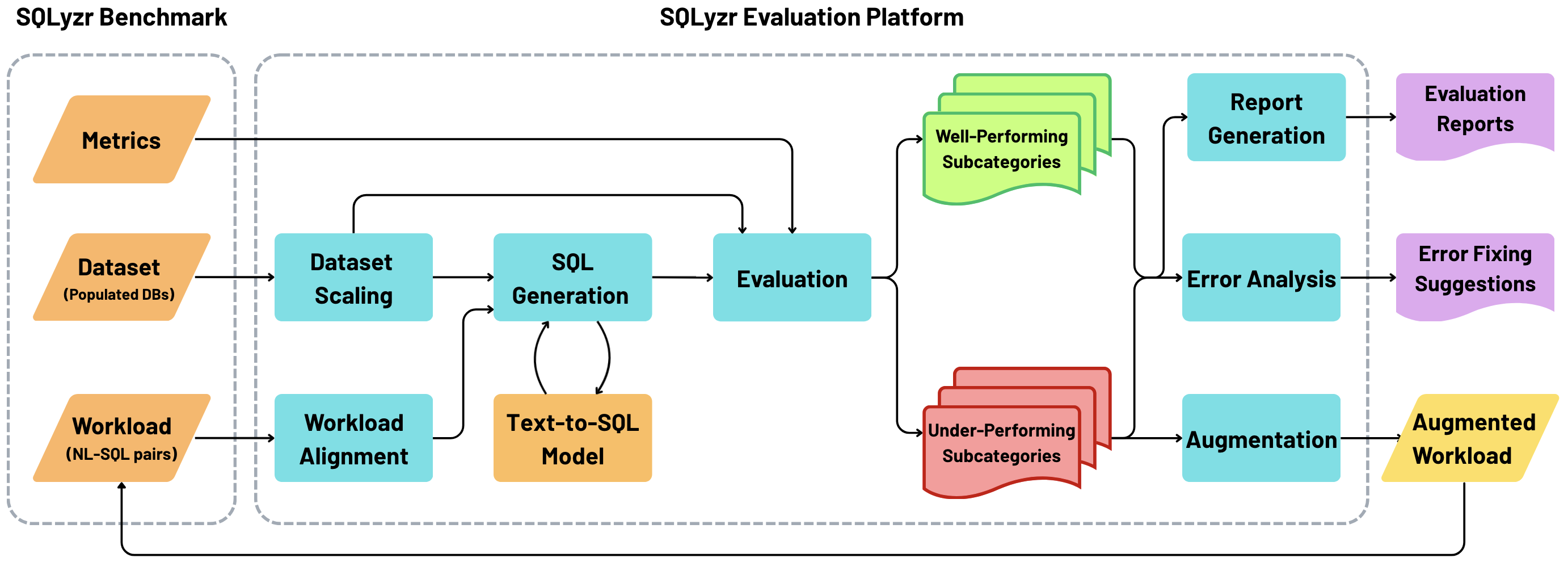}
    \caption{Overview of \name}
    \label{fig:overview}
\end{figure}

\subsection{\name\ Benchmark Specification}
\label{sec:benchmark}
The \name\ benchmark consists of three main components.

\signpost{\Wl} 
The workload is a collection of natural language questions paired with SQL queries used to evaluate \tss s. The SQL queries are treated as \gold\ and are manually annotated to ensure correctness.

\name\ introduces a comprehensive taxonomy of SQL queries based on their structural characteristics, such as nesting patterns or the presence of specific SQL constructs. This taxonomy classifies queries into six categories and 36 subcategories, ordered by increasing complexity, reflecting the progression of SQL concepts in standard database textbooks \cite{Silberschatz:2019aa}. This classification enables fine-grained evaluation by reporting scores at both category and subcategory levels, allowing users to identify which types of queries are more challenging for a \tss. It also supports the computation of complexity-based metrics and targeted workload augmentation, as described in \sref{sec:framework}.

The \wl\ includes 20{,}979 \w s, with approximately 11\% allocated for training and the remainder for evaluation. The training split is used to provide demonstration examples for in-context learning \cite{Brown:2020aa}, while the evaluation split is used exclusively for benchmarking. The \name\ \wl\ is constructed from existing Text-to-SQL benchmarks, including \spider~\cite{spider}, \bird~\cite{bird}, and \beaver~\cite{beaver}, capturing a diverse set of query types and structural complexities. 

\name\ also provides a subset of the \wl\ aligned with the empirical query distribution in SQLShare \cite{sigmod16_Jain:2016aa}, enabling evaluation that better reflects real-world SQL usage patterns. The query distribution is defined based on the frequency of queries across categories in the SQL taxonomy.

\signpost{\Ds}
The dataset consists of populated databases. Each \w\ in the \wl\ is associated with a database on which both generated and \gold\ queries are executed, and their results are compared to measure correctness and efficiency. The \name\ \ds\ is composed of databases from \spider, \bird, and \beaver, comprising 286 databases across SQLite and MySQL RDBMSs.

\signpost{Evaluation Metrics}\label{sec:metrics}
\name\ uses a set of evaluation metrics to assess \tss s that go beyond correctness, which is typically used as the sole metric in existing benchmarks. In addition to correctness, \name\ captures other important aspects of generated queries, such as efficiency, structural complexity, and generation cost.

\textbf{Execution Accuracy (EA):} Measures correctness by comparing the execution results of generated and \gold\ queries on a single database instance.

\textbf{Exact Match (EM):} Compares the structural components of generated and \gold\ queries following the Spider benchmark definition \cite{spider}. To support complex queries, we implemented an AST-based comparison that is more robust than the string-based approach originally provided in the Spider benchmark\footnote{\url{https://github.com/taoyds/spider}}, which is limited to Spider's relatively simple queries.

\textbf{Complexity Consistency (CC):} Measures whether the generated query introduces unnecessary structural complexity compared to the \gold\ query, e.g., through an unnecessary join. A query is considered consistent if its category is not more complex than the category of the \gold\ query.

\textbf{Execution Time Consistency (ETC):} Evaluates whether the generated query executes within an acceptable relative time overhead compared to the \gold\ query, capturing the execution efficiency of the generated query.

\textbf{Token Usage (TU):} Measures the number of tokens used by the language model to process input and generate the query. Since we focus on LLM-based \tss s, this metric provides an estimate of generation cost.

\subsection{\name\ Evaluation Platform \& Methodology}
\label{sec:framework}
\name\ provides a configurable platform and methodology for evaluating \tss s. An overview of \name\ is illustrated in Figure~\ref{fig:overview}. The evaluation begins by selecting a \wl, a \ds, and one or more \tss s, along with a subset of evaluation metrics defined in \sref{sec:metrics}. \name\ then executes a multi-stage pipeline that produces detailed evaluation reports and diagnostic insights.

\signpost{Dataset Scaling}
As an optional preprocessing step, \name\ supports database scaling by generating synthetic data, enabling evaluation under large-scale conditions. \name\ leverages the SDV~\cite{sdv} framework to train a generative model on each database and synthesize additional rows that preserve the statistical properties of the original data. These rows are then inserted into the databases to scale them to the desired sizes.

\signpost{Workload Alignment}
As another preprocessing stage, \name\ aligns the workload with a target query distribution. Alignment with empirical data, such as SQLShare, enables evaluations that better reflect real-world SQL usage patterns. However, \name\ is not limited to SQLShare, and users can align workloads with any target distribution. While this demonstration includes a specific instance of this functionality, full support for aligning workloads with custom distributions is available in the codebase.

\signpost{SQL Generation}
As the first evaluation stage, \name\ invokes the selected \tss\ to generate a SQL query for each natural language question in the \wl. The model is treated as a black-box component that receives a question and the corresponding database schema and produces a SQL query.

\begin{figure*}
\centering

\begin{subfigure}[t]{0.24\linewidth}
    \centering
    \includegraphics[width=\linewidth]{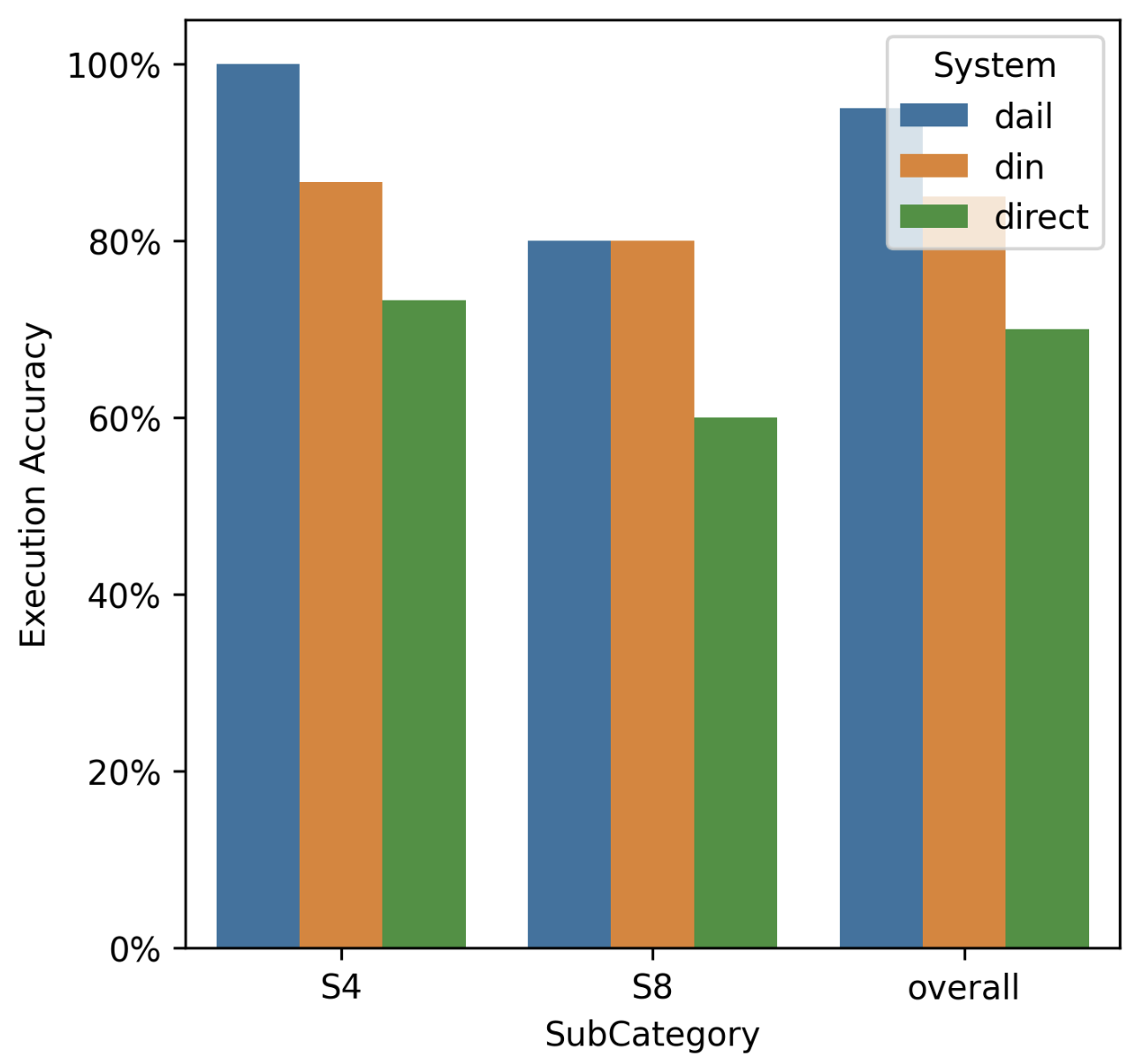}
    \caption{{\footnotesize Model comparison}}
    \label{fig:plots:systems}
\end{subfigure}
\hfill
\begin{subfigure}[t]{0.24\linewidth}
    \centering
    \includegraphics[width=\linewidth]{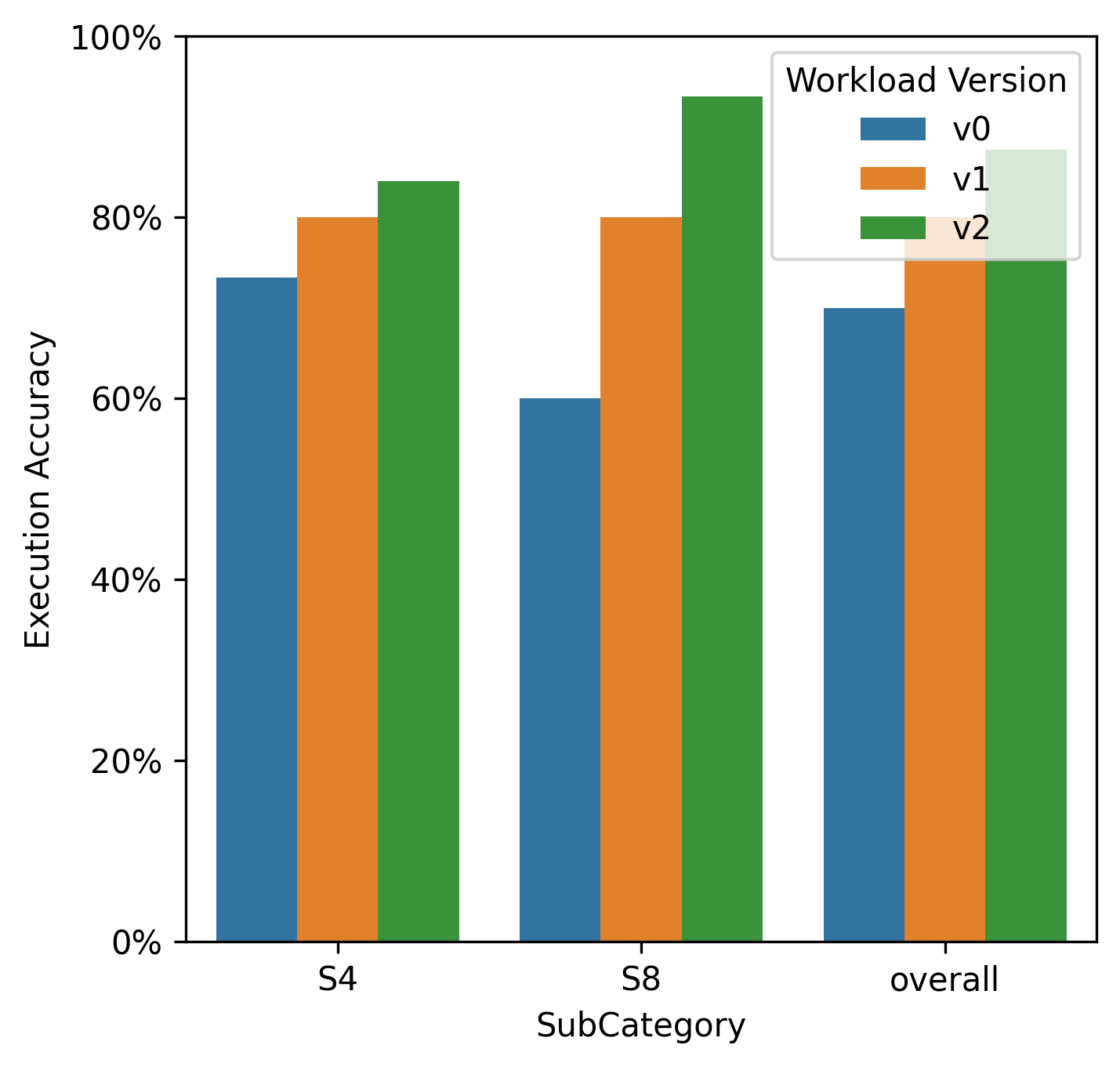}
    \caption{{\footnotesize \Wl\ comparison (fixed model)}}
    \label{fig:plots:workloads}
\end{subfigure}
\hfill
\begin{subfigure}[t]{0.24\linewidth}
    \centering
    \includegraphics[width=\linewidth]{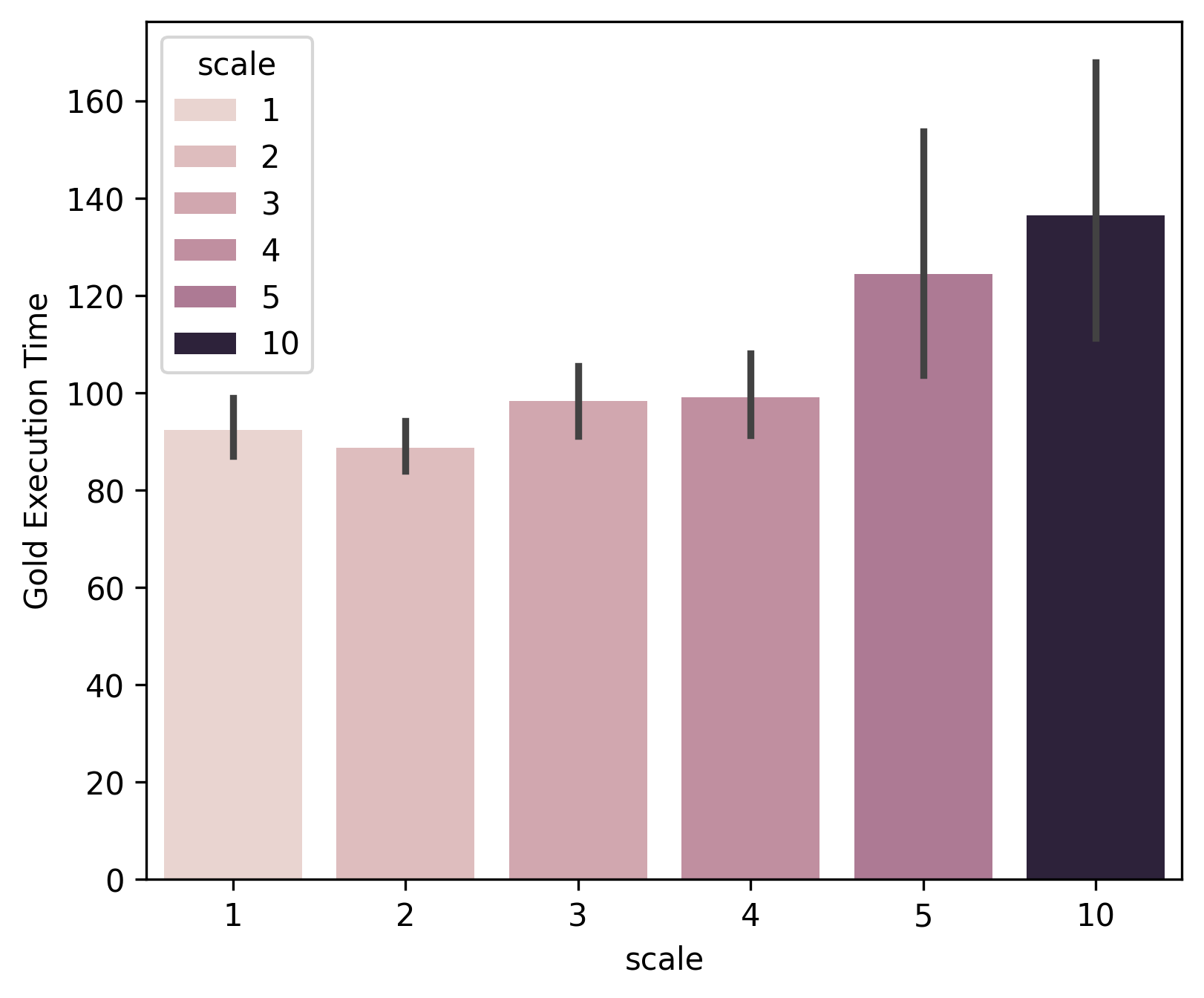}
    \caption{{\footnotesize \Ds\ scaling}}
    \label{fig:plots:scaling}
\end{subfigure}
\hfill
\begin{subfigure}[t]{0.24\linewidth}
    \centering
    \includegraphics[width=\linewidth]{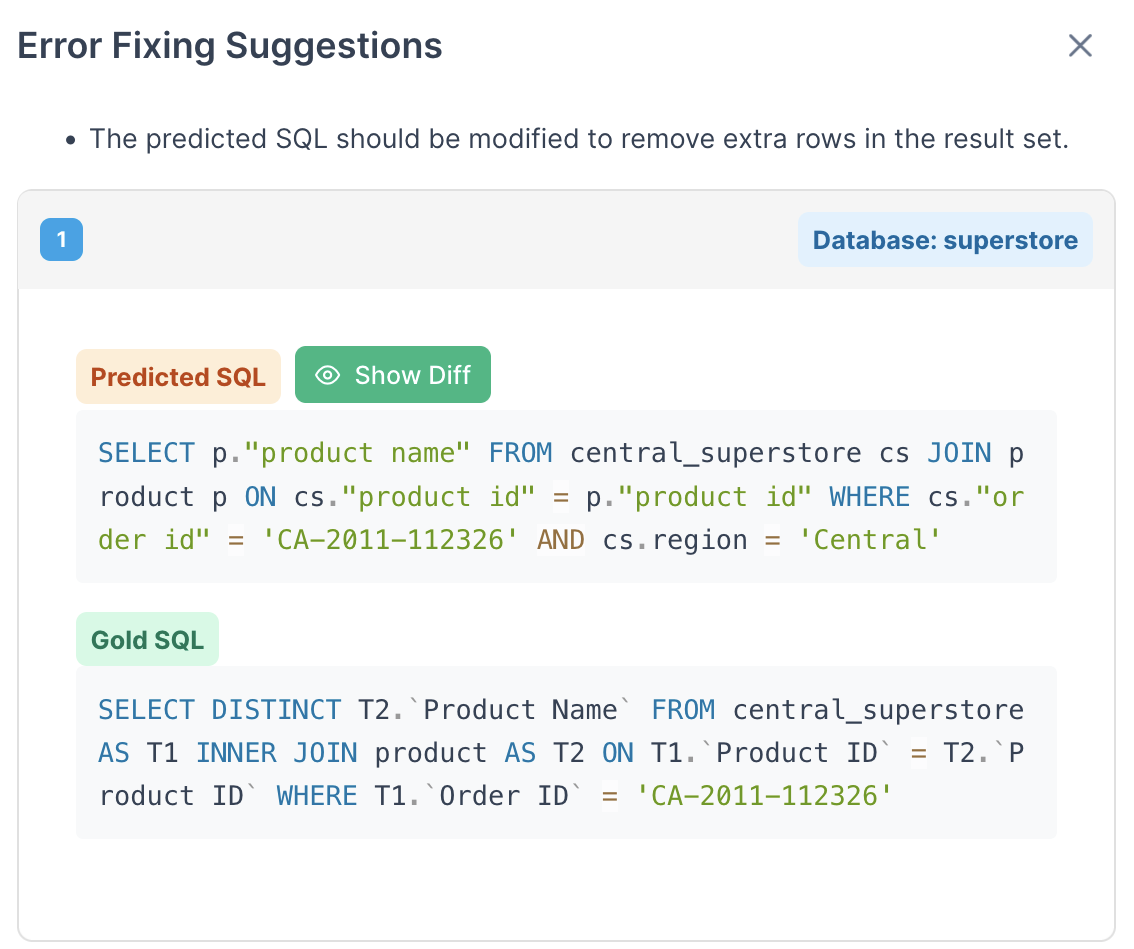}
    \caption{{\footnotesize Error fixing suggestions}}
    \label{fig:plots:repair}
\end{subfigure}

\caption{Example evaluation plots and error analysis results produced by \name}
\label{fig:plots}
\end{figure*}

\signpost{Evaluation}
In the next stage, generated queries are evaluated against the \gold\ queries using the specified metrics. This evaluation captures multiple aspects of model performance, including not only correctness but also efficiency, structural complexity, and generation cost. To support fine-grained analysis, \name\ assigns each query to a category and \sub\ based on the defined SQL taxonomy. Category assignment is implemented using abstract syntax tree (AST) traversal, which enables the identification of complex patterns in the queries.

\signpost{Analysis and Reporting}
Following evaluation, \name\ produces reports at multiple levels of granularity. Results are presented both as aggregate scores and at the level of categories and \subs, enabling detailed comparison across models. Figure~\ref{fig:plots:systems} shows an example execution accuracy plot, where both overall and \sub-level scores are reported. For each metric, \name\ also generates workload comparison plots that capture the effect of iterative augmentation (Figure~\ref{fig:plots:workloads}) and scaling plots that illustrate how scores change as database size increases (Figure~\ref{fig:plots:scaling}).

\signpost{Error Analysis}
This stage provides analysis to support the diagnosis of incorrectly generated queries. Due to the ambiguity of natural language questions, a single question may have multiple valid SQL representations. Therefore, execution accuracy can be overly strict for measuring correctness, as even minor deviations from the \gold\ query, such as different column ordering, may cause a generated query to be marked as incorrect. In this stage, \name\ identifies a subset of such cases by applying a set of transformations to generated queries and their execution results. If a transformed version of a generated query produces the same result as the \gold\ query, the corresponding transformations are reported as potential fixes. Figure~\ref{fig:plots:repair} shows an example of such repair suggestions.

\signpost{Workload Augmentation}
This stage is performed after an initial round of evaluation. Based on the reported performance, \name\ identifies underperforming \subs\ using a user-specified score threshold and generates additional \tsp\ pairs within those \subs, targeting model weaknesses. These new \w s are added to the \wl, emphasizing challenging query types and supporting iterative model improvement. This process enables repeated re-evaluation of improved models and helps uncover new weaknesses, allowing \name\ to function as an adaptive test suite rather than a one-time benchmark. Figure~\ref{fig:plots:workloads} shows an example workload comparison plot illustrating the effect of augmentation across workload versions.

\subsubsection{Implementation Details of \name}
\label{sec:extra}
This section describes implementation details and design choices that enhance the usability and efficiency of \name\ for benchmarking.

\signpost{Batch Mode}
\name\ supports both synchronous and asynchronous batch API calls for interacting with LLMs. While synchronous mode enables fast experimentation with immediate feedback, batch mode supports large-scale evaluations and reduces costs by up to 50\%. \name\ provides a unified interface for LLM interaction, allowing users to switch between modes at runtime.

\signpost{Optimizations}
\name\ leverages parallelization, asynchronous execution, and caching to accelerate evaluations. Users can also adjust runtime parameters to customize these optimizations and maximize their effectiveness.

\signpost{Configurability and Modularity}
\name\ provides various configuration options for customizing the evaluation. In addition to selecting the \tss s, users can specify parameters such as the number of evaluation iterations, the LLM used for SQL generation, and its temperature. Moreover, the platform is designed with a modular architecture that supports extensions without modifying core components. User-defined \tss s can be integrated by implementing a simple software interface. Users can also incorporate custom \wl s and \ds s beyond those already specified in the benchmark (\sref{sec:benchmark}).

\signpost{Usability}
\name\ provides both a graphical user interface (GUI) and a command-line interface (CLI) for convenient interaction with the platform. This demonstration presents the \name\ GUI.

\signpost{Database Engine Compatibility}
The current version of \name\ supports SQLite and MySQL engines. However, \name\ is not limited to these engines, and users can easily extend the support to additional engines or dialects by implementing a simple driver.

\section{Interactive Demonstration}
\label{sec:demo}
This demonstration provides interactive scenarios for evaluating \tss s using \name\ and highlights key capabilities of the platform, including fine-grained evaluation, workload augmentation, and dataset scaling. 

For demonstration purposes, we provide three small samples of \name's \wl\ with increasing difficulty levels, each containing 20 \w s. We include two state-of-the-art \tss s, \din~\cite{din} and \dail~\cite{dail}, along with a simple baseline, \textit{Direct-LLM}, which generates SQL queries from natural language questions using a simple LLM prompt. The LLM used for query generation can be selected by the user through the GUI.

\begin{figure}[!t]
    \centering
    \includegraphics[width=\linewidth]{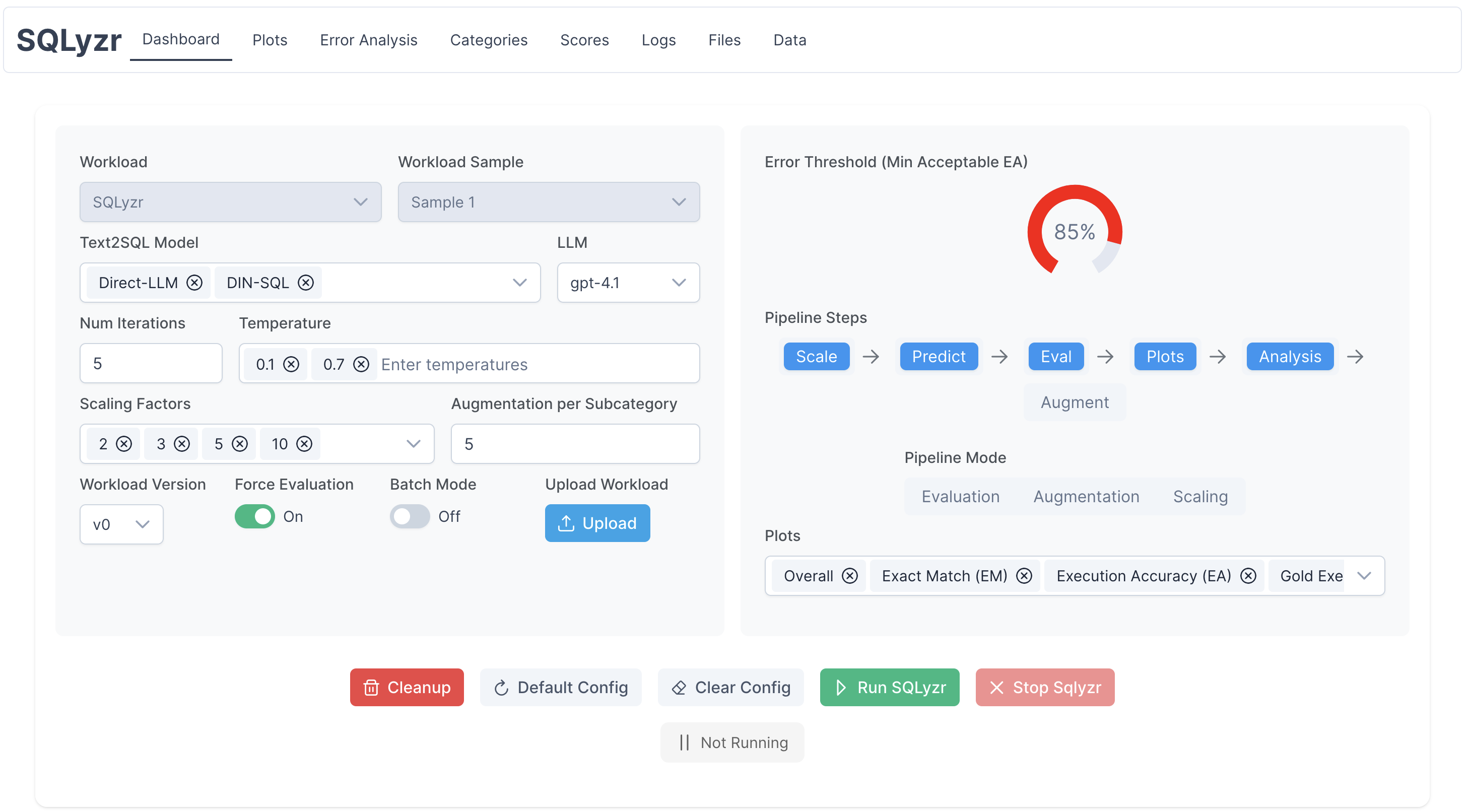}
    \caption{\name\ \emph{Dashboard} for configuring evaluation and controlling pipeline execution}
    \label{fig:dashboard}
\end{figure}

The demonstration begins with users configuring the evaluation through the main \emph{Dashboard} of \name\ (Figure~\ref{fig:dashboard}). Users then start the evaluation by clicking the \emph{Run SQLyzr} button. Upon execution, \name\ produces a set of plots that visualize fine-grained evaluation results, enabling detailed analysis and comparison.

\signpost{Scenario 1: \Mdl s Comparison and Diagnosis}
In this scenario, users evaluate and compare multiple \mdl s using one of the sample \wl s. After selecting models and configuring parameters such as the number of iterations, the LLM choice, and the temperature, users start the evaluation and monitor its progress through the \emph{Logs} panel. Upon completion, the \emph{Plots} panel presents results at both aggregate and fine-grained levels. These fine-grained plots reveal differences in model behaviour that are not visible through aggregate metrics alone, allowing users to identify model strengths and weaknesses across query types (Figure~\ref{fig:plots:systems}). The \emph{Error Analysis} panel further highlights incorrect but fixable queries and suggests potential fixes, helping users to understand the causes of model errors and diagnose failures more effectively (Figure~\ref{fig:plots:repair}).

\signpost{Scenario 2: Iterative \Wl\ Augmentation}
This scenario demonstrates how \name\ supports adaptive evaluation through workload augmentation. Starting from an initial \wl, users identify underperforming \subs\ based on the evaluation results and specify a score threshold to guide augmentation. \name\ then generates new \w s targeting these \subs\ and extends the \wl\ accordingly. By repeating this process, users construct progressively refined workloads that emphasize more challenging queries for the evaluated model. Users can then inspect evaluation plots that compare metrics across different versions of the augmented \wl s (Figure~\ref{fig:plots:workloads}).

\signpost{Scenario 3: Evaluation under \Ds\ Scaling}
The final scenario demonstrates \name's \ds\ scaling capability using the same \wl\ from the augmentation scenario. Users begin by selecting desired \emph{Scaling Factors} through the \emph{Dashboard} and executing the evaluation pipeline via the \emph{Run SQLyzr} button. In the first stage, synthetic data is generated and inserted into the databases, after which \name\ evaluates the \mdl\ on the scaled data. Following execution, \name\ produces plots showing how evaluation scores evolve as database size increases (Figure~\ref{fig:plots:scaling}). This enables users to assess model performance under conditions that more closely resemble real-world settings. 

\section{Conclusion}
In this work, we present \name, a comprehensive benchmark and evaluation platform for \tss s. Through an interactive demonstration, users explore \name's capabilities via its GUI. We highlight several features of \name\ that address key limitations of existing benchmarks, including fine-grained evaluation, \wl\ augmentation, and \ds\ scaling. We envision that \name\ facilitates the evaluation and iterative refinement of \tss s by moving beyond static benchmarking toward an adaptive and development-oriented evaluation framework.

\bibliographystyle{ACM-Reference-Format}
\bibliography{refs}

\end{document}